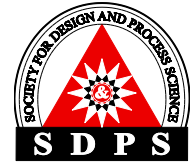

# Digital Engineering Transformation with Trustworthy AI towards Industry 4.0: Emerging Paradigm Shifts

Jingwei Huang
Department of Engineering Management and Systems Engineering, Old Dominion University
Norfolk, VA 23529, USA
j2huang@odu.edu

**Abstract** Digital engineering transformation is a crucial process for the engineering paradigm shifts in the fourth industrial revolution (4IR), and artificial intelligence (AI) is a critical enabling technology in digital engineering transformation. This article discusses the following research questions: What are the fundamental changes in the 4IR? More specifically, what are the fundamental changes in engineering? What is digital engineering? What are the main uncertainties there? What is trustworthy AI? Why is it important today? What are emerging engineering paradigm shifts in the 4IR? What is the relationship between the data-intensive paradigm and digital engineering transformation? What should we do for digitalization? From investigating the pattern of industrial revolutions, this article argues that ubiquitous machine intelligence (uMI) is the defining power brought by the 4IR. Digitalization is a condition to leverage ubiquitous machine intelligence. Digital engineering transformation towards Industry 4.0 has three essential building blocks: digitalization of engineering, leveraging ubiquitous machine intelligence, and building digital trust and security. The engineering design community at large is facing an excellent opportunity to bring the new capabilities of ubiquitous machine intelligence and trustworthy AI principles, as well as digital trust, together in various engineering systems design to ensure the trustworthiness of systems in Industry 4.0.

**Keywords** Industry 4.0, Fourth Industrial Revolution, digitalization, digital transformation, digital engineering, digital engineering transformation, paradigm shift, data science, digital technologies, uncertainties, trustworthiness, trustworthy AI systems

## 1. Introduction

The convergence of disruptive digital technologies, such as artificial intelligence (AI), Internet of Things (IoT), cloud computing, high-speed wireless communication, and blockchain, among others, has driven the world into a new wave of a technology revolution, known as the "fourth industrial revolution" (4IR) (Schwab, 2017), also known as the "digital revolution" for the central role of digitalization in the 4IR. In 2011, Germany first launched their Industrie 4.0 initiative aiming at leveraging Cyber-Physical-Systems (CPS) to build smart factories and strengthen their leadership in manufacturing. The US launched a national Advanced Manufacturing program in 2014 and declared manufacturing to be a national priority. Recent Five-year Plans of China since the 2010s actively promote critical technologies towards 4IR. Industry 4.0 (I4.0) is a term frequently used as an interchangeable term for 4IR; however, they have some subtle differences in their focuses. In this article, we use the 4IR to address the revolutionary change in paradigms and tools, and we use Industry 4.0 to address the targeted industrial systems (with manufacturing in the core as the fundamental component) in the 4IR. Briefly, Industry 4.0 is the target or goal, and the 4IR is the movement or process toward that goal.

A defining feature of the fourth industrial revolution is digitalization. Associated with this feature, a pervasive and profound digital transformation is ongoing in almost every aspect of human society globally. Digital Engineering is the digital transformation in the field of engineering, and digital engineering transformation is at the center of 4IR. In 2018, the US DoD launched their Digital Engineering Strategy (US DoD, 2018; Zimmerman, Gilbert, & Salvatore, 2019). This strategy requires using and sharing formal models and digital data across engineering lifecycle and organizational boundaries through a trusted "authoritative source of truth". This move will impact the US defense industry and propagate to other industry sectors and change how engineering is conducted. Digital engineering transformation is about engineering paradigm shifts in the 4IR, and AI is a critical enabling technology in the digital engineering transformation.

AI has achieved remarkable progress in recent years. At the same time, the disruptive impacts and fast growth of applications of AI systems (particularly ML) also raise broad concerns about an AI system's reliability, safety, security, privacy-preserving, fairness (or bias-free), explainability, traceability, transparency, and accountability, among other qualities. Research on those concerns has been ongoing under the umbrella of "Trustworthy AI" (AAAI Presidential Panel on Long-Term AI Future, 2009; EU AI HLEG, 2019; H. Liu et al., 2021; NSF, 2019; Stone et al., 2016; Wing, 2020). Trustworthy AI has two primary aspects: societal effects (ethics) and technical performance (dependability). The ethics of AI concerns the long-term impacts of AI on humans and human society. The central principle is to use AI for good or for purposes beneficial to humans. Dependability concerns the competency of AI systems on technical matters. Given the critical role of AI in the 4IR, the recent concerns about trustworthy AI also propagate into Industry 4.0 systems. The AI community has been striving for many decades to push the boundary of machine intelligence. The engineering community will bring AI technologies and ethical principles together to deliver trustworthy cyber-physical-social smart systems to human society.

This article will discuss the following research questions:
(1) What are the fundamental changes brought by the fourth industrial revolution?
(2) More specifically, what are the fundamental changes in engineering? What is digital engineering?
(3) What are the main uncertainties associated with the 4IR? What is trustworthy AI?
(4) What are emerging engineering paradigm shifts in the 4IR?

The contents of this article are organized as follows. Section 2 discusses the answer to the first question by investigating the patterns of the four industrial revolutions, focusing on the ongoing fourth revolution. Section 3 continues with the second question to discuss digital engineering transformation in the 4IR. Section 4 discusses the uncertainties in the 4IR and trustworthy AI. Section 5 discusses some open issues and focuses on the emerging engineering paradigm shifts. Finally, section 6 concludes the article.

## 2. The Fourth Industrial Revolution

What are the fundamental changes (including paradigm shifts and their profound impacts) brought by the fourth industrial revolution? Let us logically examine the patterns that emerged in industrial revolutions historically from the perspectives of technological triggers, paradigm shifts, and profound impacts. As illustrated in Figure 1, the fundamental changes of both the first industrial revolution (1IR) and the second industrial revolution (2IR) are the provision of energy (physical power) for industrial production. The fundamental changes of the third industrial revolution (3IR) and the fourth industrial revolution (4IR) are the provision of machine intelligence (brainpower) to explore and leverage information power in industries and human society. Both 1IR and 3IR are fundamental for inventing revolutionary new tools (the steam engine and the computer) to deliver physical and information power, respectively. Both 2IR and 4IR scale up the delivery of new powers (energy and information power) initiated in 1IR and 3IR, respectively.

The 1IR is characterized as "mechanization." The invention of the steam engine triggered the 1IR. Steam engines lead to the industrial operations paradigm shift from human manual labor to the broad use of machines in production. A representative example is frequently referred to as the first mechanical loom.

The profound impact of the first industrial revolution is that the using steam-powered machines to replace human hands in production activities significantly improved the efficiency of production and released humans from heavy manual labor for more creative activities.

The 2IR is characterized as "electrification." The invention and use of electric power, together with the invention of assembly lines, triggered the second industrial revolution. The use of electricity makes power (energy) usage easy, flexible, free from the space-constraints, and easy to scale up or down. Without electricity, the idea of assembly lines could not have been realized. With easy access to electric power, the invention of assembly lines leads to skill specialization-based mass production. This change marks the industrial operations paradigm shift from early craft production and job production to much more productive mass production and the associated product standardization. The profound impacts of the second industrial revolution are mass production and flexible and scalable access to power (energy).

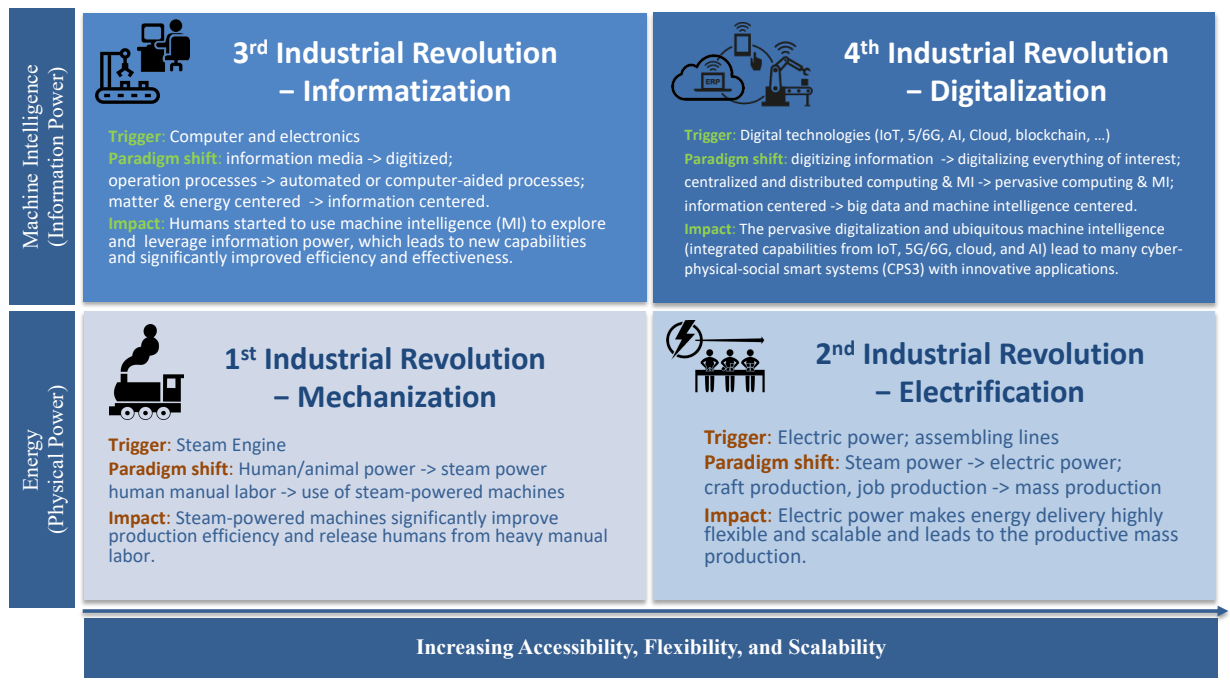

Figure 1. Four waves of Industrial Revolutions and their profound impacts. Every wave was triggered by disruptive technologies and followed by paradigm shifts in industrial operations.

The 3IR can be characterized as "informatization." The invention and development of programmable general-purpose digital computers triggered the 3IR. The computer is the most powerful information processing tool in human history. The availability of this critical tool allows humans to explore and leverage the power of information, along with using matter and energy. The paradigm shift brought by the 3IR can be viewed in three aspects. First, with the broad applications of computers in control, such as the adoption of PLCs, DCS, and SCADA, many human-operated processes became automated or computer-aided processes. Secondly, information media was digitized (or computerized). Research reveals that 94% of information media became digital in 2007 (Hilbert & López, 2011). The digital forms of information allow information to be stored and read with computers. This machine-readable (and later machine-processable) property makes information flow largely automated and fast. The advantages of digital information make information a decisive success factor beyond other factors during the time of the 3IR. This change leads to the third aspect of the paradigm shift - the business process transition from material, energy, and product centered to information centered. The fundamental impact of 3IR is the beginning of using machine intelligence to explore and leverage the power of information, which leads to new capabilities and significantly improved efficiency and effectiveness in industrial production and other human activities.

For this reason, the 3IR is also called the "information revolution," and the period is called the "information age." Since the start of the 3IR, many digital technologies, such as artificial intelligence, industrial robots, portable and then mobile computers, the Internet, wireless networks, mobile devices, and others, were developed and profoundly changed human society regarding how we work and live already. Those digital technologies prelude the next wave of technological revolution.

The 4IR can be characterized as "digitalization." The convergence of transformative and disruptive digital technologies, such as IoT, high-speed wireless communication, cloud computing, AI&ML, and blockchain, among others, trigged the 4IR.

IoT, in a broader perspective, also named as "Internet of All Things" or "Internet of Everything," has four pillars: Internet of Things, Internet of Information, Internet of People, and Internet of Services (Kirkpatrick et al., 2013). In the future internet infrastructure, the physical things of interest (such as various machines, devices, buildings, roads, and other structures, even things of interest in the natural world), the abstract things (things in the digital world, such as information contents), people, organizations, and various processes of interest (examples include: service processes, or engineering processes, even social or natural processes of interest, such as climate change, environmental change, living things status change, pandemic monitoring, disease infection tracing, and others), anything of interest are interconnected, through future internet, operating with supports from 5G/6G wireless communication networks, cloud computing, and AI. In the rest of this paper, we refer term *IoT* in this broader sense of the *Future Internet* or *Internet of Everything*.

IoT connects the physical world with the cyber world to form an integrated cyber-physical smart environment. With IoT through sensors and actuators, smart things in the physical world can be remotely located, monitored, and even controlled through the Internet. With mobile devices supported with 5G/6G wireless networks, people will have ubiquitous access to the Internet, smart things in the physical world, and the cloud services needed to handle those smart things. More importantly, people themselves become nodes of this Internet. This interconnected world, facilitated by IoT, 5G/6G, cloud computing, and AI&ML, is leading to numerous Cyber-Physical-Social Smart Systems (CPS3) (Huang, Seck, & Gheorghe, 2016), such as autonomous vehicles, smart factories, smart stores, smart cities, and others. A necessary condition to develop those smart systems is to digitalize things of interest to make them uniquely identifiable and machine-understandable; then, we can leverage AI&ML to exploit the unprecedented richness of information brought by the digital and connected world.

The 4IR with those disruptive digital technologies is "leading to unprecedented paradigm shifts in the economy, business, society, and individually" (Schwab, 2017). The emerging paradigm shifts in the 4IR can be viewed from several related perspectives as follows.
(1) Shift from digitizing information to digitalizing everything of interest, including artifacts (both abstract and physical), natural things, organizations, and processes (such as production and/or service processes, monitored natural processes, and others).
(2) Shift from centralized and distributed provisioning of computing services and machine intelligence to pervasive provisioning of computing services and machine intelligence.
(3) Shift from information-centered processes (developed in 3IR) to big data and machine intelligence-centered processes.

About the first shift, different from the 3IR, where information was digitized, in the 4IR, not only information (abstract artifacts) but also physical artifacts and natural living things of interest, as well as processes of interest, could be digitalized. Everything of interest, even humans, could have their digital counterparts (or "digital presence" as termed by (Schwab, 2017)). Such as, some devices or products will have their digital twins, and some artifacts will have their digital augmentations (the associated metadata, e.g., provenance). In addition to artifacts, some processes of interest (no matter about production, service, or some natural processes such as climate change or regional ecosystem change, among others) could also be digitalized. This pervasive digitalization could be enabled by semantic web technology (a branch of AI) (Berners-Lee, Hall, Hendler, Shadbolt, & Weitzner, 2006; Berners-Lee, Hendler, & Lassila, 2001), RFID, IoT, and blockchain.

The second shift can be seen from comparing how computing service was delivered before the 4IR. From many people using a single large computer, each person using a computer and then a laptop, till each computer was able to connect to the Internet with a wall jack, computing service was delivered in a "centralized", "decentralized", and then "distributed" approach. In the 4IR, we can use computers and interact with the world everywhere.

About thirty years ago, Mark Weiser and Nicholas Negroponte had a debate at MIT Media Lab. Negroponte predicted that AI was going to be the leader for the next wave of computing, but Weiser argued it would be Ubiquitous Computing, characterized as "invisible" and "connection" (Weiser, 1996). Weiser's vision, which can be found in a vivid illustration in his essay "Open House" (Weiser, 1996), is one of the original ideas about the Internet of Things. Interestingly, IoT and AI together become the major drivers for the 4IR. IoT extends the "nerves" of the Internet into the physical world, and AI is the "brain" of smart things interacting with the digitalized connected world. In addition to IoT and AI, wireless networks and cloud computing are critical contributors to the pervasive provisioning of computing and machine intelligence.

First, of course, high-speed wireless communication is an essential infrastructure to facilitate the mobility of communication, computing, sensing, and delivery of all types of digital services without or significantly reducing the physical space constraints. Secondly, cloud computing enables ubiquitous on-demand network access to a shared pool of computing resources, thus meeting scalable and elastic computing needs (Armbrust et al., 2009; NIST, 2011). Cloud computing is essential for handling the big data produced in the digital world and for the intensive computing required by AI applications, no matter where we are.

This article uses term "*ubiquitous machine intelligence*" (uMI) to refer to the integrated capabilities enabled by the convergence of IoT, 5G/6G, AI&ML, and cloud computing. Interestingly, the role *ubiquitous machine intelligence* plays in meeting the needs of computing and machine intelligence is very similar to the role of electricity in 2IR, where electricity enables the mobility of power delivery needed in assembly lines and mass production. Here, *ubiquitous machine intelligence* makes computing services and machine intelligence pervasively accessible and provides the essential technology for various smart systems. For this reason, *ubiquitous machine intelligence* is regarded as the most fundamental change brought by the 4IR.

The third shift can be viewed through the following comparison. In the Industry 3.0, information flows and material flows were still separated; thus, information is long-delayed. In the 4IR, the pervasive digitalization and connectedness make the material and human flows associated with information flows; thus, real-time information about the material and human flows can be used for decision-making. There will be unprecedented rich information about the things of interest for decision making. Consequently, a decisive success factor in the era of 4IR will be the capability of leveraging ubiquitous machine intelligence to collect and extract information, discover knowledge from big data, and make smart decisions in the digital smart and connected environment (Huang, 2017).

Finally, regarding the profound impact of the 4IR, similar to the 2IR that makes power (energy) easily accessible and leads to mass production, the 4IR makes information power (computing and machine intelligence) pervasively accessible and leads to various CPS3 with innovative applications. More specifically, in the 4IR, the pervasive digitalization and pervasive connectedness make data (minerals of information) available in an unprecedented scale and make computing services and machine intelligence (brainpower or information power) ubiquitously accessible and scale up. Furthermore, they are triggering and accommodating numerous innovative applications. Those disruptive technologies and their innovative applications are "fundamentally changing the way we live, work, and relate to one another." They lead to "the transformation of entire systems, across (and within) countries, companies, industries and society as a whole" (Schwab, 2017).

## 3. Digital engineering transformation

As discussed in the last section, digitalization is the central theme of the 4IR. As such, pervasive digital transformations are ongoing in almost every aspect of human society. Given the essential role of engineering in industries, the digitalization of engineering is at the core of 4IR. Digital engineering is the digitalized engineering, or the targeted digital form of engineering to be realized by the digitalization of engineering towards Industry 4.0.

The US Department of Defense (DoD) defines "digital engineering" as "an integrated digital approach that uses the authoritative source of system data and models as a continuum across disciplines to support lifecycle activities from concept through disposal" in their "Digital Engineering Strategy" (US DoD, 2018; Zimmerman et al., 2019). As illustrated in Figure 2, the strategy targets five goals: (1) formalize the development, integration, and use of models, leading to a continuous end-to-end digital representation of the system of interest; (2) provide an enduring authoritative source of truth to share and exchange digital models, data, and other digital artifacts across boundaries of organizations and the engineering lifecycle; (3) incorporate technological innovation to improve the engineering practice; (4) establish a supporting infrastructure and environments; (5) transform culture and workforce to adopt and support digital engineering. Paper (Zimmerman et al., 2019) provides a comprehensive view of the US DoD's efforts for digital engineering transformation. The implementation of this strategy will significantly change how engineering practice is conducted in the US DoD enterprise and propagate to industries in a broader range.

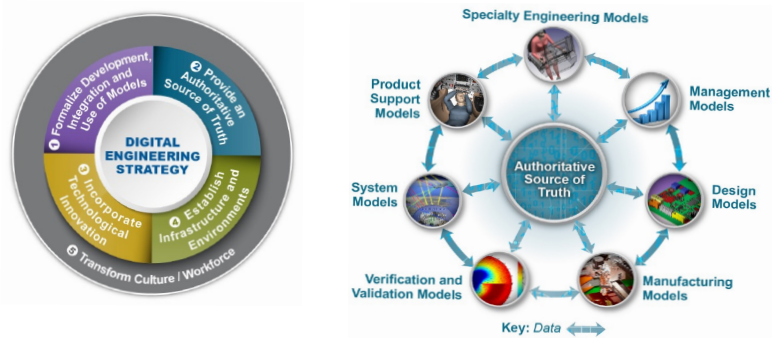

Figure 2. The US DoD Digital Engineering Strategy. Picture from (US DoD, 2018) (Fig.3&4)

From the perspective of engineering processes, the transformation from conventional engineering to digital engineering is illustrated in Figure 3. In the picture, from a general view, an engineering process is illustrated as a process model with inputs (left), outputs (right), enablers (bottom), and control (top). Digitalizing engineering artifacts, engineering processes, and engineering enterprises is the foundation for digital engineering transformation. Without digitalized artifacts, the applications of emerging digital technologies will be very limited, maybe in an ad hoc manner, or may need human or ad hoc devices as intermediate agents. As a result, the level of automation and autonomous functioning will stay at a slightly higher level than in Industry 3.0. On the other hand, with digitalized engineering artifacts and processes, we can broadly leverage machine intelligence and other digital technologies to conduct engineering in a digital, smart, and connected environment; thus achieving a new level of engineering, which is more effective and efficient as well as more trustworthy. Some advantages of digital engineering include: sharable knowledge and data across the engineering life cycle, increased explainability and transparency of engineering processes and products, fast integration of distributed engineering resources for a given mission, increased traceability and accountability, increased product traceability along supply chains, and quick adaption to the changing environment, among others.

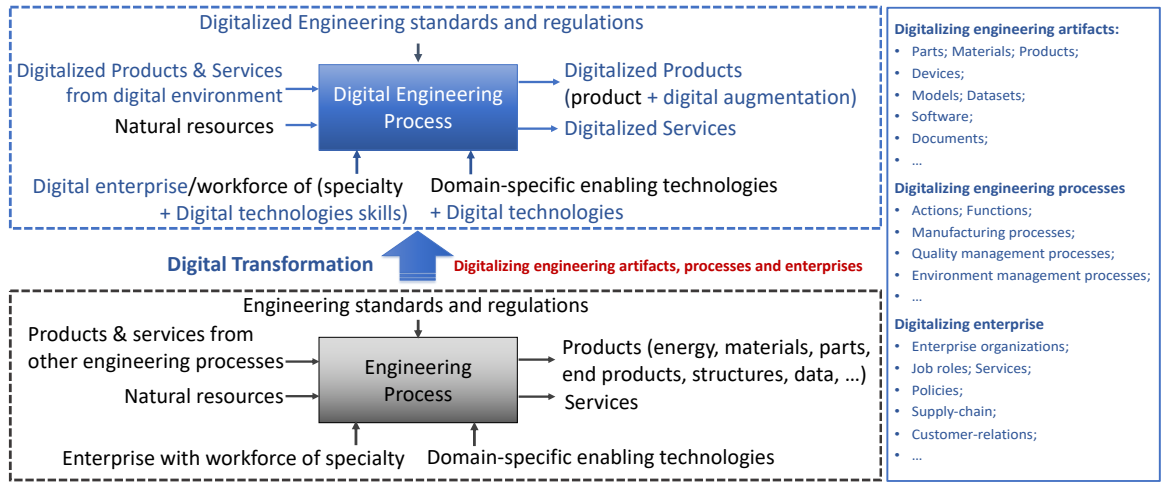

Figure 3. Digital engineering transformation from a process perspective

|  | Concept Stage | Development Stage | Production Stage | Utilization/support Stage | Retirement Stage |
|---|---|---|---|---|---|
| Model creation. | Inputs:<br>- Digital artifacts (DAs) in operating environment;<br>- Relevant data and models from downstream stages. | Inputs:<br>- Digital artifacts from both upstream and downstream;<br>- Digital artifacts of external systems | Inputs:<br>- Digital artifacts from both upstream and downstream;<br>- and from production environment. | Inputs:<br>- Digital artifacts from both upstream and downstream;<br>- and from operating environment. | Inputs:<br>- Digital artifacts from upstream;<br>- and from natural environment. |
| Model learning | Apply AI&ML for model building from big data coming from upstream and downstream engineering stages and environment. | | | | |
| Model integration | Interaction between digital models for both SysCon and systems in operating environment. | Interaction between digital models for both system components and external systems. | Interaction between digital models for both the system and systems in production environment. | Interaction between digital models for both the system and external systems in operating environment. | Interaction between digital models for the sys component and DAs in natural environment. |
| Model curation | Create model of models; maintain metadata for models; maintain model provenance; model update and propagation to downstream. | | | | |
| Model sharing & use | Across lifecycle activities,; across the boundaries of disciplines and organizations | | | | |
| Model Trustworthiness | Centralized standardization; decentralized standardization and mappings; distributed evolutionary fine-grained convergence; model trustworthiness; model repeatability; Access Control; digital artifact intellectual property protection, … | | | | |

Figure 4. Model perspective: operations in the digital engineering lifecycle (Huang et al., 2020)

In digital engineering as shown in Figure 3, the inputs, outputs, enablers, and control are different from the conventional ones; correspondingly, the digital engineering operations will also significantly differ from the conventional ones. In the digital engineering transformation, along with the fast-paced digital technologies, "computational thinking" (Wing, 2006, 2012) is critical to the new engineering paradigms. Knowledge and skills in applied computing and relevant digital technologies have become necessary components in workforce development towards digital engineering. Figure 4 briefly describes operations in the digital engineering lifecycle from a model perspective. Some further discussion of digital engineering can be found in (Huang et al., 2020).

Finally, AI is a critical enabling technology to advance digital engineering (Huang, Beling, Freeman, & Zeng, 2021). In addition to the broad applications of Machine Learning for modelling and prediction in engineering, another AI branch, knowledge representation and reasoning, particularly semantics technology, is a fundamental technology to support digitalization. For example, ontologies are used in MBSE (Madni & Sievers, 2018). AI also plays a critical role in the logical formalization in building and operating digital trust & security systems (Huang, 2018). We will discuss further in subsection 5.3.

## 4. Harness uncertainties in the 4IR – A key: trustworthy AI systems

The disruptive digital technologies in the 4IR not only offer transformative opportunities but also bring in uncertainties and have triggered broad concerns and research on systems trustworthiness. The focused concerns of trustworthiness have been shifting and expanding to cover new issues that pop up in the new technologies used in each generation of engineering systems. This section discusses the trustworthiness concerns of Industry 4.0 systems, starting from examining the evolution of the trustworthiness of engineering systems, then focusing on trustworthy AI systems – the pivotal component of Industry 4.0 systems.

Before we move to these topics, let us discuss the elementary concepts of trust first. We use trust explicitly or implicitly, even unaware of using it, in our daily life and work. However, what is trust exactly? There are many definitions of trust (Walterbusch, Gräuler, & Teuteberg, 2014). Our working definition of trust used in this article is as follows (Huang & Fox, 2006) - ***Trust*** *is a mental state comprising (1) expectancy: the trustor expects a specific behaviour of the trustee, such as providing true information or effectively performing cooperative actions; (2) belief: the trustor believes that the expected behaviour will occur, based on the evidence of the trustee's competence, good intention, and integrity; (3) willingness to take the risk: the trustor is willing to take the risk for (or be vulnerable to) that belief*. In brief,

> *Trust   =   Expectation +*
> *Belief in that expectation +*
> *Willingness to take the risk for that belief.*

Trust can be regarded as a relation between a "trustor" (trusting party) and a "trustee" (trusted party) with respect to the properties of something provided/offered/conducted by the trustee. For example, a passenger (trustor) trusts an autonomous car (trustee) regarding the safety of driving on the streets. What is to be trusted is out of the control of the trustor. From the studies of trust, the factors leading to trust include the trustee's ability, good intention (or benevolence, as some researchers called it) towards the trustor, and integrity (reflected by the principles and values guiding behaviours)(Mayer, Davis, & Schoorman, 1995). The roles of ability and good intention in the concept of trust are apparent. Integrity is the need for trust judgment with respect to the predictable behaviours of the trustee in situations that are out of watching by the trustor.

According to Simon (Simon, 1997), a decision-making process in the real world is limited by "bounded rationality," i.e., the "rational choice that takes into account the cognitive limitations of the decision maker - limitations of both knowledge and computational capacity." In the real world, because we only have limited information/knowledge, limited computing capacity, and limited time available for decision-making, a decision has to be made partly based on bounded rational calculation and partly based on trust. As Luhmann addressed (Luhmann, 1979), trust, as "a basic fact of social life," functions as a "reduction of complexity." Without trust, social interactions, including the applications of technologies in engineering and business, will be extremely difficult. Because of trust, we do not need to verify everything before a decision making or action. Trust mechanisms (Huang & Nicol, 2013) play a critical role in harnessing the disruptive digital technologies toward trustworthy utilizations of them.

### 4.1. Trustworthiness of systems

Trustworthiness of a system is defined in (Schneider, 1999) as "assurance that a system deserves to be trusted—that it will perform as expected despite environmental disruptions, human and operator error, hostile attacks, and design and implementation errors." In short (Avizienis, Laprie, Randell, & Landwehr, 2004), systems trustworthiness is the "assurance that a system will perform as expected." Trustworthiness is a relative concept and is dependent on what is expected. *Trustworthiness is often characterized as a collection of the expected properties of a system to be trusted,* as illustrated in Figure 5. The "assurance" can be achieved by various trust mechanisms implemented in engineering systems and in societal systems such as audits, certifications, regulations, and law enforcement.

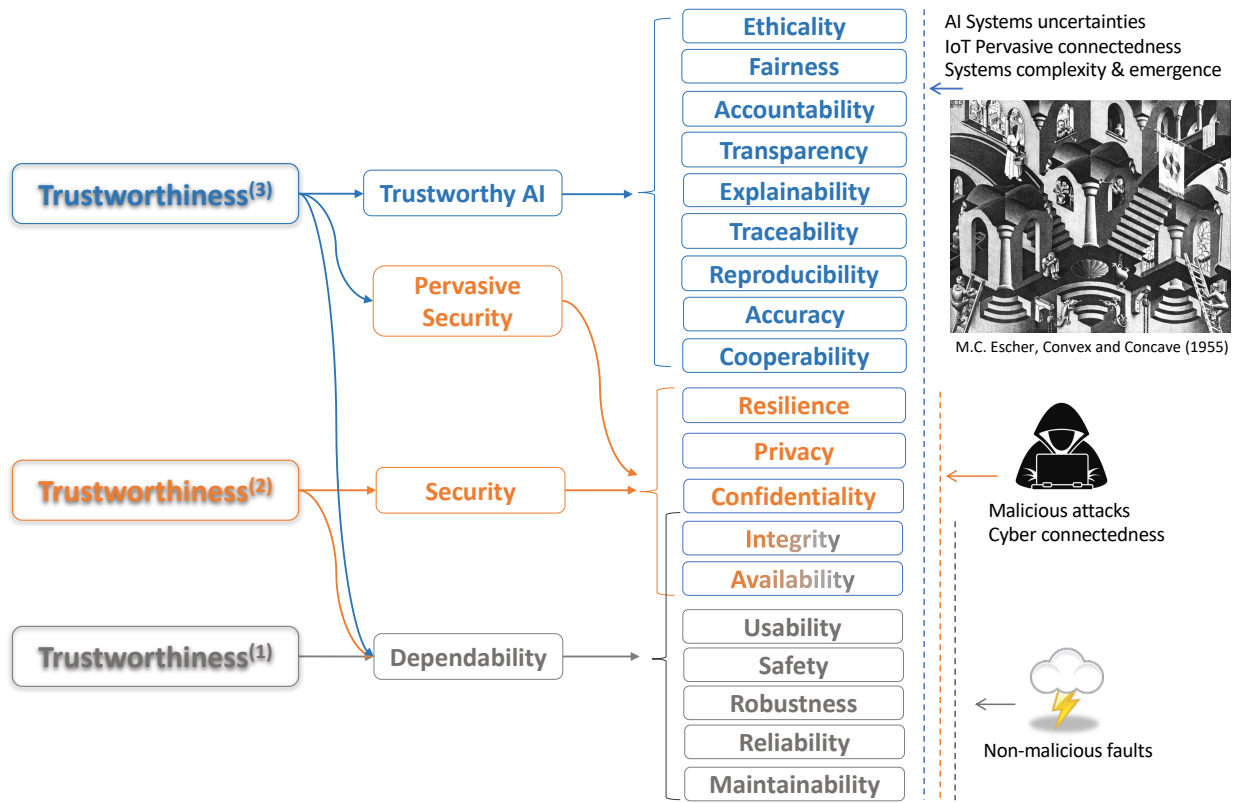

Figure 5. The evolving trustworthiness properties as the evolution of systems with technologies and the sources of concerns

In the following, we discuss the expected properties of trustworthiness of different generations of engineering systems and the sources of the concerns or threats to trustworthiness. Traditional systems are typically stand-alone systems (a physical system, a computer system, or a system embedded with computer systems) or just connected logically and internally within an organization rather than the Internet. As shown in the bottom part of Figure 5, the trustworthiness properties concerned with those traditional systems are mainly reliability, maintainability, availability, integrity, robustness, safety, and usability. Those properties together are referred to as dependability. The sources of concerns or the threats to the trustworthiness properties are mainly non-malicious faults caused by (1) errors introduced in systems design and development; (2) errors introduced in manufacturing and maintenance; (3) errors happened in system operations; (4) environmental disruptions, including natural disasters, environmental pollution, and gradual but irreversible environment change associated with climate change.

In the middle part of Figure 5, enabled by Internet and wireless mobile communication, the networked industrial systems evolve to cyber-physical systems (CPS) (Lee, 2008; NSF, 2008; Rajkumar, Lee, Sha, & Stankovic, 2010), which use networked computing components to monitor, coordinate, control, and integrate physical systems and processes. The connectedness of systems naturally led to great challenges with respect to security and privacy, and correspondingly they become the central theme of systems trustworthiness (Avizienis et al., 2004; Schneider, 1999). The threats to security and privacy are obviously from various malicious attacks and incompetent security defence.

As illustrated in the upper part of Figure 5, trustworthiness for the systems in the 4IR involves concerns with trustworthy AI, pervasive security, and technical dependability. The issues in an AI system could trigger security issues and systems dependability issues; of course, security issues could also trigger systems dependability issues. The sources causing the systems trustworthiness issues are in three folds:

(1) The pervasive security issues caused by the pervasive connectedness facilitated by IoT: the pervasive connectedness facilitated by IoT broadly extends the security and privacy concerns to every corner of the world where humans work and live. Without assured technologies and societal mechanisms, the Internet of Things can turn into "the Internet of Risky Things" (S. Smith, 2017), even the Internet of dangerous things. The threats to security and privacy in 4IR include malicious attacks and incompetent security defence as before but with a much larger attack/exposure surface. Additionally, the threats also include certain relevant societal mechanisms. For example, as Smith pointed out (S. Smith, 2017), some smart devices will last longer than their manufacturers and suppliers, and who will keep updating the systems or their embedded computing components? Without updates, a digital system will be vulnerable in the complex digital operating environment.
(2) The uncertainties caused by AI systems: the fast growth and broad applications of AI systems or embedded AI components in various systems have triggered much concern in two aspects: technical competence (such as reliability and security) and ethical concern (such as transparency, bias towards groups of people, human-machine relations, and others). The next subsection will discuss this aspect.
(3) systems complexity & emergence - the aforementioned two primary two factors together bring in high uncertainties and complexities into the systems in the 4IR. Complex systems typically exhibit strong emergence (unexpected emergent system behaviour and properties that were not anticipated in design and development, e.g., the US-Canada Northeast Blackout of 2003) (Adcock, Jackson, Fairley, Singer, & Hybertson, 2021).

### 4.2. Trustworthy AI

In the last decade, AI achieved remarkable advances. Just to name a few, AlphaGo (Silver et al., 2016, 2018), using deep neural networks-based reinforcement learning, won the world number one player of Go game, which is regarded as the most challenging board game for computers to win. AlphaGo's achievement marks a new milestone of machine intelligence towards superhuman intelligence on specific tasks. Although it is still a pilot project and has no broader adoption yet, Waymo One, the commercial taxi service operating with level 4 autonomous vehicles, has been offered on the streets in a city (Waymo One, 2020). This service marks a new level of integrated intelligent capabilities of an AI system on a type of complex tasks which previously only humans could do. Another remarkable advance is Generative Adversarial Networks (GANs) (Creswell et al., 2018; Goodfellow et al., 2014, 2020). A GAN consists of a pair of deep neural networks, a generator, and a discriminator; together, they can generate realistic images and other contents as requested (see examples at https://thispersondoesnotexist.com/). GANs lead to many potential applications but also open the door for deepfake.

For many decades, the AI community's focus has been on extending machine intelligence capabilities. Although the advances of AI today are still domain-specific (or narrow AI) rather than general AI, the remarkable achievements, the disruptive impacts, and the fast growth of applications of AI systems triggered concerns and research on trustworthy AI systems. The concerns are mainly from two perspectives. (1) Ethical AI, focusing on social effects, mainly ethical considerations, essentially, the profound effects AI systems bring to humans, human groups, and human society, and advocating using AI to benefit humans as a central principle. (2) Reliable AI, focusing on the technical performance, or dependability, concerning the competency of AI systems on technical matters. There is a broad range of expected properties of trustworthiness in AI systems.

Ethical AI is an emerging interdisciplinary research field, gathering together researchers from a wide range of areas, including computer science, philosophy, sociology, anthropology, public policy, law, and others. Some thought leaders and researchers conducted pioneering work in the field. The Association for the Advancement of Artificial Intelligence (AAAI) organized a panel of leading AI researchers that conducted "Asilomar Study of Long-Term AI Futures" in 2008. The study explored a wide range of topics on societal impacts and guidance of AI research (AAAI Presidential Panel on Long-Term AI Future, 2009). The study assessed concerns and perspectives about disruptive outcomes of superhuman intelligence, explored ethical and legal issues associated with autonomous systems, and identified near-term AI research challenges and opportunities, including enhancing people's privacy, enhancing human-AI collaboration and interaction, making ML and reasoning transparent to people, and preventing using AI for malevolent purpose. Continuing the effort to guide AI research for good, the AI100 2016 report (Stone et al., 2016) addressed that AI research is "shifting from simply building systems that are intelligent to building intelligent systems that are human-aware and trustworthy." The report collectively presents the view and prospects of a panel of experts on the opportunities and challenges of AI and policy recommendations in eight selected AI application domains targeting people's lives in a typical North American city in 2030. They addressed challenges regarding how to build safe and reliable systems, gain public trust concerning safety, security, and privacy, make AI systems behave ethically and overcome bias, and how AI systems smoothly interact with humans. In the long term, ethical AI concerns the fundamental relations between AI systems and humans. There are concerns about "technological singularity" or "intelligence explosion," partially reflected by the dystopia depicted in fiction. Experts in the AI field believe those radical outcomes remain fictional and are not immediate threats; well, some thought leaders suggest "avoid strong assumptions regarding upper limits on future AI capabilities" (Future of Life Institute, 2017). Given AI's profound impacts on human society, it is necessary to review AI systems' purposes, usages, and impacts and create principles and regulations for guiding AI research and development to avoid harming humans and human society. The 23 Asilomar principles (Future of Life Institute, 2017) and the EU HELG ethical AI guide (EU AI HLEG, 2019) reflect a broad range of concerns about ethical issues of AI systems. EU AI HELG defined trustworthy AI as three components: lawful AI, ethical AI, and robust AI. The guide proposed four ethical principles: (1) Respect for human autonomy; (2) Prevention of harm; (3) Fairness; (4) Explicability, covering transparency, auditability, traceability, and explainability. On the ground of these four principles, the guide further proposed a list of seven requirements: (1) Human agency and oversight; (2) Technical robustness and safety, including security, accuracy, reliability, and reproducibility; (3) Privacy and data governance; (4) Transparency, including traceability, explainability, and communication; (5) Diversity, non-discrimination and fairness; (6) Societal and environmental wellbeing; (7) Accountability. The EU guide also discussed technical and non-technical methods to implement the requirements and proposed a list of questions used to assess the trustworthiness of AI systems.

The ethics of AI studies what is right/wrong regarding the purpose and usages of AI, based on the profound effects AI systems bring to humans and human society. The central principle of ethical AI is about using AI to benefit humans and human society. In the upper part of Figure 5, ethicality is about the extent to which an AI system complies with this principle.

As addressed by the AI100 2021 report (Littman et al., 2021), "AI systems and humans have complementary strengths;" thus, "combined, they can accomplish more than either alone." Technically, it remains a challenge regarding how to team up humans and AI systems effectively. For the fundamental ethical principles of trustworthy AI (EU AI HLEG, 2019), no doubt human-AI teaming is the right direction to go, not only for maximizing capability and performance but also for ethical consideration. *Cooperability* is about the extent to which an AI system facilitates and supports human-machine teaming for complex problem-solving, including the channels or methods to enhance interactions and collaborations between humans and machines. *Cooperability* covers c*ontrollability; the latter* is about the ability of an AI system that allows humans to monitor and control the system.

*Fairness* is about whether an AI system fairly treats people of different groups regarding race, gender, age, cultural background, and others. Fairness received much attention in recent years when AI systems started being used for some life-changing scenarios, such as hiring decisions, financial credit evaluation, and judicial decisions (Mehrabi, et al., 2021). Fairness is a highly challenging topic for complex human societal reasons. Technically, fairness can be treated as bias-free. The recent deep learning progress reflected by GANs can be used to create real-like samples for balanced data in ML.

*Accountability* is the availability and integrity of the identity of an entity that performed an operation in the AI system of concern. In simple, it is who (human operators or autonomous components in an AI system) did what and when and the responsibility for that. In the scenarios of a security incident, an accident, an error, or a system failure, accountability helps to identify the causes, make responsibility clear, and avoid future mistakes.

*Transparency* is about the extent to which how the AI system operates is transparent to various stakeholders, such as operators, business partners, auditors, and users. Transparency is a basis for ensuring ethicality, fairness, and privacy and facilitating controllability and cooperability.

*Explainability* is the ability to explain the outcomes and processes of an AI system to humans. The major challenge is that in ML, neural networks have low-level coding for feature representation which is inherently hard to be explained in high-level knowledge. A very large number of parameters and complex structures of deep neural networks make the interpretability further harder.

*Traceability* is about the ability to collect and document the provenance of the data used and produced, the models used and trained, and the operating processes of an AI system. Traceability is the basis for transparency and supports explainability.

*Reproducibility* is about whether a model instance can be rebuilt with the same AI algorithm and data and whether an experiment or, more generally, a process running with an AI model can be reproduced. Reproducibility is essential to science and engineering.

Reliable AI reflects people's concern about systems' technical performance when more and more AI systems are used or embedded in a large engineering system in the 4IR. Naturally, this aspect of concern brings the focused attention partially back to more classical trustworthiness properties of engineering systems but with a focus on AI systems or AI components and their impact on the larger systems. This article uses the term "dependability" to cover all expected properties (for trustworthiness) on the technical performance aspect. The concept of dependability here is broader than what was defined in (Avizienis et al., 2004). *Reliability* is the probability of a system functioning without failure for a given period of time. *Maintainability* is about how easy to make a system maintain healthy, updatable, and upgradeable. *Resilience* is about the ability of a system to restore it to a working state when it is damaged in situations such as natural disaster events or cyber-attacks. *Safety* is about how safe to humans a system is. The broad use of AI components in engineering systems makes safety a significant concern. *Accuracy* is the measure of errors made by a model or system; in the context of ML, it is critical to have high accuracy for new data beyond the data used for training the model. In Systems Engineering, robustness is defined as "the degree to which a system or component can function correctly in the presence of invalid inputs or stressful environmental conditions"(ISO/IEC/IEEE, 2010). In ML, *robustness* is about the stable outcomes in the presence of perturbations in inputs and could be measured with sensitivity. In deep learning, lack of robustness is a critical cause for the possible deepfake by GANs. The general meaning of robustness is similar to resilience. In robustness, the perturbations are on a small scale. Usability is how easy a system is to be used, and usability is beyond human-machine interaction. Poor usability can lead to failures in many other properties in modern systems, including security and safety.

*Security* is defined as the well-known definition of the CIA triad: *Confidentiality* (Prevention of unauthorized access to the protected resources or disclosure of the protected information), *Integrity* (absence of unauthorized alterations), and *Availability* (Readiness for correct services for authorized users). Some other interesting security properties can be defined on top of the CIA triad. For example, *authenticity* is the integrity of information content and its provenance (origin). Obviously, in the digital and connected environment, failures in security have broad impacts and can compromise other trustworthiness properties. Privacy is another big concern in AI. Since data is the fuel for AI systems, privacy concerns about whether the data gathering, holding, processing, usage, sharing, and governance respect people's privacy.

As illustrated by Figure 5, the trustworthiness of AI systems is the new collection of concerns when the systems evolve into Industry 4.0 systems. Interestingly, on the other hand, the digitalization of engineering artifacts, processes, and enterprises in the 4IR could support achieving the expected trustworthiness properties of AI systems.

In digital engineering transformation, it is an excellent opportunity for the engineering design community at large to bring the new capabilities of AI and the trustworthy AI principles together in various engineering systems design for human society to leverage the power of AI and at the same time to avoid or minimize the potential negative impacts.

## 5. Open issues and discussion

In previous sections, we have discussed the characteristics of 4IR and their fundamental impacts, digital engineering transformation - the manifestation of 4IR in engineering, and trustworthy AI - the leading technology in the digital transformation for I4.0. This section continues the discussion on some open issues in the direction of 4IR.

### 5.1. What is the pattern of paradigm shifts?

As discussed in section 2, digitalization in the 4IR is leading to paradigm shifts in almost every aspect of our society. What are the emerging new engineering paradigms? Engineering uses scientific principles to design, build, and operate engineering systems for solving real-world problems or meeting humans' needs in a specific field, thus essentially sharing the paradigms in scientific research. A general pattern of scientific paradigm shifts can be illustrated as the following Figure 6.

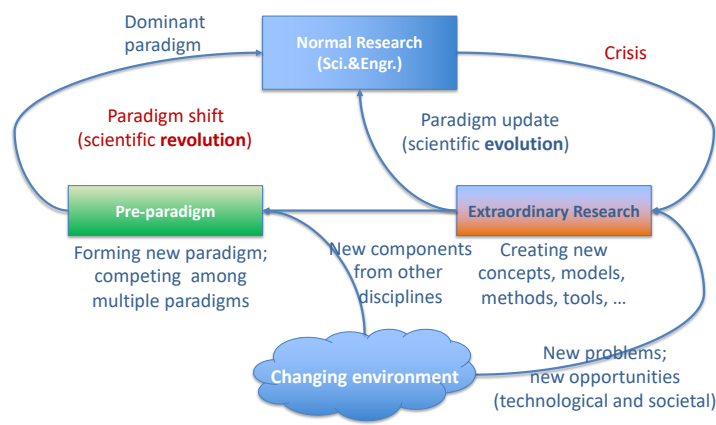

Figure 6. A view of paradigm shifts, based on Kuhn's structure of scientific revolutions (Kuhn, 1962).

According to Kuhn's structure of scientific revolutions (Kuhn, 1962), in the period of "normal research" or "normal science," a scientific community uses a dominant paradigm to conduct research and produce the main body of knowledge in that field. Then, the dominant paradigm may have crises for its deficiencies or limitations facing the new observations and/or new problems coming from the changing environment. To battle the crises and meet the new need(s), the disciple enters a period of "extraordinary research" (as named by Kuhn). In this period, new concepts, models, tools, methods, among others, will be created. If the new disciplinary components are incremental and can be integrated into the current paradigm, the paradigm will be updated, thus being a scientific evolution. Otherwise, the new components developed in "extraordinary research," possibly together with the components from other disciplines, will contribute to the development of a new paradigm in a period of "pre-paradigm" (again, named by Kohn). In the "pre-paradigm" period, one or multiple paradigms will be formed and compete. Finally, the most accepted paradigm(s) will become the discipline's new dominant paradigm(s). This "paradigm shift" (Kohn) is a scientific revolution. After the transformation, the discipline enters again "normal research" period and starts a new life cycle.

The above discussed pattern is revealed in the context of scientific research. It appears to be general, as we treat a scientific paradigm as a knowledge system that evolves in the environment.

### 5.2. What is the fourth paradigm of scientific research?

Jim Gray had a vision that scientific research is shifting to the fourth paradigm (data-intensive paradigm) after the empirical, theoretical, and computational paradigms (Gray, 2007; Hey, Tansley, & Tolle, 2009), as illustrated in Figure 7. The empirical paradigm is characterized as finding and describing patterns based on the observations of real-world phenomena, and the theoretical paradigm is characterized as building theories on mathematical models. The computational paradigm uses computer simulations to tackle the difficulty when the theoretical models become too complex to derive and prove propositions.

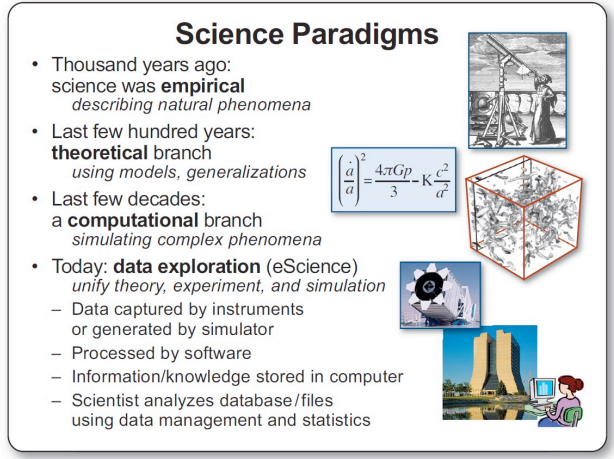

Figure 7. Four Paradigms in Scientific Research, Slide from (Gray, 2007; Hey et al., 2009)

The fourth paradigm reflects the scientific research in the new research environment with the big data produced from instruments and simulations, high demands of computing for knowledge discovery from big data, and emerging demands of data publishing and research work reproducibility. In the fourth paradigm, scientific research significantly focuses on knowledge discovery from big data and reproducible models and datasets sharing.

It is essential to point out that a new paradigm does not mean entirely replacing existing paradigms. Instead, a new paradigm is established on the ground of existing paradigms. A new paradigm represents the focus shift and the new approach to address new problems in a new environment. Theoretical models are built on the ground of observations, and computational simulations are based on theoretical models. The fourth paradigm enhances and unifies the empirical, theoretical, and computational paradigms, on the ground of an unprecedented big data environment and the new technology for knowledge discovery from big data.

Inspired by Jim Gray's fourth paradigm, with the fast growth of big data in the last decade, data science has emerged as a new discipline. A NIST publication (NIST Big Data Public Working Group, 2019) pointed out that "data science is the fourth paradigm of science." However, it has been heavily focused on knowledge discovery from data. From the library and information science perspective, data curation and knowledge curation are among data science topics. Still, the needs emerging from the fourth paradigm are beyond just traditional curation and go much further into knowledge and data sharing with traceability, explainability, accountability, reproducibility, and interoperability.

By the fourth paradigm, data-intensive science/engineering will emerge in various disciplines of science/engineering. (Data-intensive engineering is the manifestation of data science in the engineering field.) In this paradigm shift, data science is the common core, shared by many domain-specific science/engineering disciplines that need knowledge discovery from big data and knowledge sharing. The combination of data science (the fourth paradigm) with each specific discipline will form many domain-specific data-intensive science/engineering disciplines.

### 5.3. What are the emerging new engineering paradigms?

"Science is the systematic description of phenomena" (Richards, 1928). Science focuses on discovering the essential laws of nature. Engineering is "the application of science to the optimum conversion of the resources of nature to the uses of humankind" (R. J. Smith, 2022). On the one hand, there is an intersection between science and engineering, as engineering includes scientific research for discovering applied knowledge for engineering purposes. From this perspective, *data-intensive engineering* is the manifestation of the data-intensive paradigm of science in the digital and connected operating environment of the 4IR.

On the other hand, engineering activities are beyond discovering new knowledge and focus more on the design, manufacturing (or construction), operating, and support of useful systems for human society, thus having much more complex interactions and relations with other systems in the operating environment, including human stakeholders. Given this complexity of engineering, the disruptive digital technologies, and the associated digital, connected, and smart environment in the 4IR, how should we conduct engineering? What are the new engineering paradigms in the 4IR?

No doubt, the *data-intensive paradigm* of scientific research is a crucial one. It includes both aspects of knowledge discovery from data and knowledge sharing (including data). The current "data science" central theme has focused more on knowledge discovery from data. While "digital engineering" in the US DoD's vision is more from the perspectives of model-based engineering and sharing of models and data. Interestingly, model-based engineering is an effort, like the development of theoretical and computational paradigms in science, to introduce models in engineering workflows, including traditionally informal activities such as requirement elicitation, requirement representation, system concept modelling, system design, among others. Certainly, modelling needs to be based on scientific knowledge as much as possible. Also, the availability and richness of data in the engineering environment and the fast progress of machine learning make it a powerful way to build models from data, leading to data-intensive engineering.

However, given the complex interactions and relations between engineering, human society, and the operating environment, the data-intensive paradigm (that focuses on scientific discovery) alone is insufficient for tackling the complexity of engineering. The need for engineering paradigm shifts in the 4IR is driven by the engineering environment, the disruptive digital technologies, the associated higher social-economic needs, and the new challenging problems, such as the trust issues of AI systems (as discussed in Section 4). In the landscape of the 4IR, the digital engineering transformation needs new concepts, models, tools, methods, theories, methodologies, technologies, and standards.

As discussed in section 2, the 4IR has three aspects of paradigm shifts: digitalizing everything of interest, provisioning ubiquitous machine intelligence, and big data and machine intelligence-centred business processes. On the other hand, digitalization and ubiquitous machine intelligence also triggered broad concerns and potential issues of trustworthiness, as discussed in section 4. To put the above-discussed pieces together, we could have a relatively clear big picture in the direction of Industry 4.0, as illustrated in Figure 8, in which we need three interdependent essential building blocks for the digital transformation of engineering and industries: (1) digitalization of engineering; (2) leveraging ubiquitous machine intelligence; (3) building digital trust and security.

Digitalization is a foundation to realize ubiquitous machine intelligence and needs to support digital trust. For this consideration, to digitalize a thing of interest (which could be an object or a process), we need the following four essential components:

(1) Digital representation of the thing of interest in a standard form with well-defined semantics to make it universally accessible by different types of machines on different platforms.
(2) A unique identifier of the thing of interest, which is a necessary component for traceability, verifiability, accountability, and explainability.
(3) Associated metadata, such as provenance, in a standard form with well-defined semantics to enable the use of digital technologies to manipulate and operate the thing automatically.
(4) Verifiable association of the unique identifier, the digital representation, and the metadata with the thing of interest. This association ensures the authenticity and integrity of the digital artifact.

For the digitalization of engineering, basically, we need (i) to digitalize engineering artifacts, engineering processes, and enterprises; (ii) enable the sharing and interoperability of digitalized artifacts across the engineering lifecycle; (iii) to develop digital model-based engineering. Virtualization is a view from the perspective of operating with digitalized artifacts and processes.

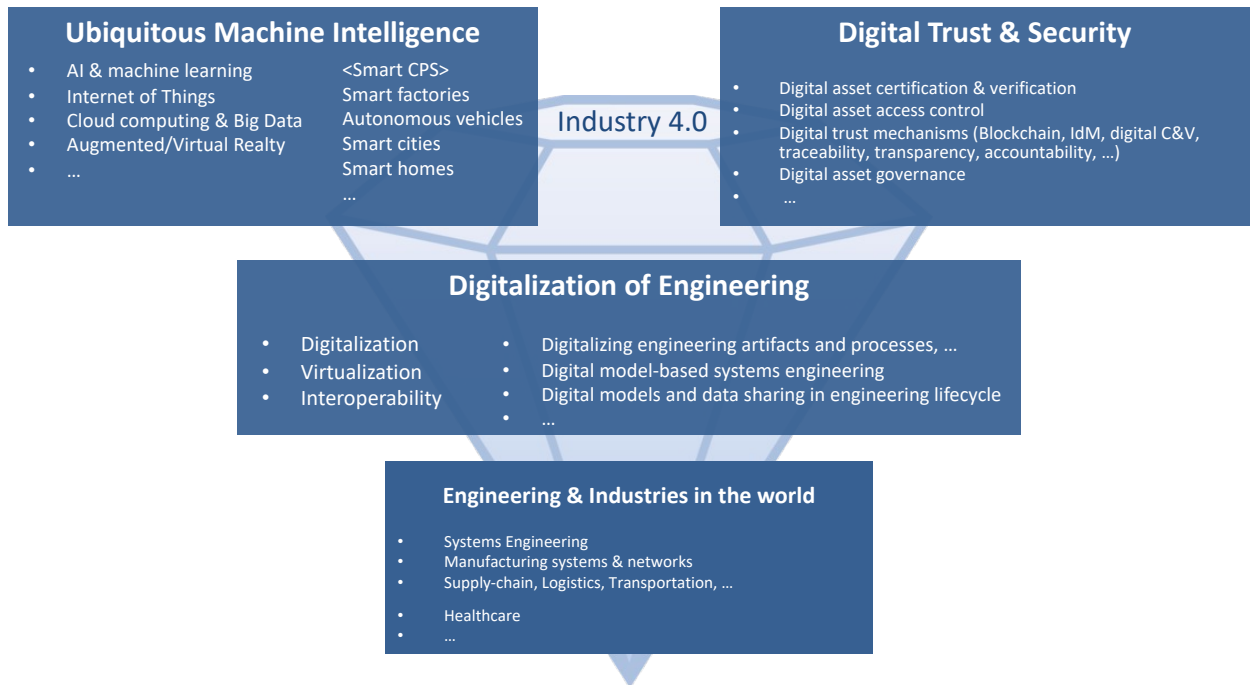

Figure 8. Three essential building blocks for digital engineering towards Industry 4.0: digitalization of engineering, leveraging ubiquitous machine intelligence, and building digital trust & security.

Digitalization needs to be realized with digital trust in design and supported by ubiquitous machine intelligence among digital technologies. Naturally, in this direction, what are enabling technologies for digitally modelling engineering artifacts, processes, and enterprises? Semantic web technology (a branch of AI) (Berners-Lee et al., 2006, 2001), RFID, IoT, and blockchain are essential. Also, directly related, intensive research on enterprise modeling and enterprise integration has been conducted since the 1990s and can be used to support model sharing across boundaries (Chen, Doumeingts, & Vernadat, 2008; Fox & Gruninger, 1998; Fox & Huang, 2005; Goranson, 2002; Vernadat, 2020). Enterprise modelling combined with digital identity management could form a basis for digitalizing engineering artifacts. Ontologies play an essential role in the digitalization of engineering with respect to artifacts, entities, organizations, various engineering activities and processes, among others (Ahmed, Kim, & Wallace, 2006; Demoly, Kim, & Horváth, 2019; Dimassi et al., 2021; Fox, Barbuceanu, Gruninger, & Lin, 1998; Gruninger & Fox, 1995; H. M. Kim, Fox, & Grüninger, 1999; K.-Y. Kim, Manley, & Yang, 2006; Sirin, Coatanéa, Yannou, & Landel, 2013).

   The realization of ubiquitous machine intelligence, the second aspect of paradigm shifts in the 4IR, is on the top of digitalization. With digitalized engineering artifacts and processes, we can apply ubiquitous machine intelligence in engineering to design, build, and support smart products, provide smart services, build trustworthy supply chains, and conduct engineering intelligently. Consider the following examples. Machine learning has been broadly applied in additive manufacturing (Z. Jin, Zhang, Demir, & Gu, 2020; Meng et al., 2020; Wang, Tan, Tor, & Lim, 2020). In the digital, smart, and connected environment, unprecedented big data provide rich information for using AI in engineering operations. Condition-based maintenance or predictive maintenance are typical scenarios where AI&ML can be leveraged (Black, Richmond, & Kolios, 2021; Carvalho et al., 2019). Engineering involves many sequential decision making; with rich information about engineering systems' status, reinforcement learning can be used to learn from simulations, experiments, routine operations, and generally experience for optimization, such as in mesh generation (Pan, Huang, Cheng, & Zeng, 2022), in manufacturing (Su, Huang, Adams, Chang, & Beling, 2022), for engineering design (Dworschak, Dietze, Wittmann, Schleich, & Wartzack, 2022). Machine learning can also be applied to even traditionally labour-intensive and time-consuming requirement elicitation process (Cheligeer et al., 2022; Mokammel et al., 2018). Essentially, engineering systems design starts with the environment (Zeng, 2004, 2015, 2020). In the new engineering environment of the 4IR, many engineering systems become cyber-physical-social smart systems, and it is a challenging issue regarding how to design such systems (Horváth, 2022; Tavčar & Horvath, 2018).

Digitalization brings to us not only advantages but also new issues, most significantly, trust and security issues with digital artifacts and those digital technologies. Some examples of those issues include how do we ensure the authenticity and integrity of digital artifacts, what are policies regarding who can access what, when to access and where to access, and how do we ensure traceability, transparency, accountability, and reproducibility, among others. We must develop proper digital mechanisms to address those issues. How could we achieve secure information sharing in digital engineering? As an example, a security policy integrating role-based access control and attribute-based access control (including security classification-based mandatory access control) (Huang, Nicol, Bobba, & Huh, 2012; X. Jin, Sandhu, & Krishnan, 2012; Servos & Osborn, 2017) can help. Digital identity management and digital trust mechanisms (Huang & Nicol, 2013), particularly blockchain (Y. Liu et al., 2020; Zheng et al., 2020), can support trust management of digital artifacts. Digital archives are critical to ensure data integrity and proof of existence (Vigil, Cabarcas, Buchmann, & Huang, 2013), and blockchain facilitates a new approach to distributed digital archives. Scientific computing integrity (Huang, 2018; Peisert, Cybenko, & Jajodia, 2015) can be further developed to support trusted engineering workflows.

Digital Engineering is an emerging form of engineering in the digital revolution. Many issues and questions there need to be researched. Just name a few, in digitalization, should the standard forms for digital representation and augmentation be supported by a centralized standardization (such as creating international standards) or a distributed evolutionary standardization (such as many ontologies competing to be standards at a fine-grained level and evolving gradually and naturally)? How do we achieve trustworthy AI systems in 4IR? What are digital trust mechanisms for digital engineering? Many trust mechanisms used in cloud service (Huang & Nicol, 2013) are applicable to digitalized products, systems, and services, while still, what new mechanisms should be introduced? Should be the sharing of digital engineering models and data in a centralized way or distributed way? There are many issues and challenges ahead as we conduct engineering in a very different new digital environment (Coatanéa, Nagarajan, Panicker, & Mokhtarian, 2022; Horváth, 2022; Huang et al., 2020).

### 5.4. What is next?

Earlier in section 2, we discussed the journey of four industrial revolutions in human history. Potentially, what is next? Given the broad and profound impacts of digital engineering transformation on human society, a farther vision will help us to develop long-lasting engineering paradigm(s) and to better design and develop digital engineering systems.

Fig. 9 illustrates the journey of industrial revolutions from passively following nature to actively exploiting nature and possibly achieving harmony between human society and nature. Let us start the discussion with the physical limits of computing. Computational power is the core of ubiquitous machine intelligence, the defining power we gain from the 4IR. However, with Moore's law reaching its limit, the chips for CPUs have hit the ceilings of size and performance, particularly power consumption. The High-Performance Computing (HPC) community has been striking hard to achieve an exascale computing system with power consumption within 20MW (Lucas et al., 2014). In the most recent Top500 List issued in June 2022, the current fastest supercomputer *frontier* is the first exascale machine with 1.1 ExaFlop/s at 21MW. What we hope to beat Moore's law are new computational technologies emerging in the horizon, such as new semiconductor materials, Quantum Computing, and DNA computing.

Data centres are the most significant contributors to today's computational power, and data centres also consume a tremendous amount of energy and have significant environmental footprints. For example, in the US, data centres consumed 1.8% of electricity in 2014 and were responsible for 0.5% of greenhouse emissions in 2018 (Siddik, Shehabi, & Marston, 2021). It has become more significant regarding how to improve data centres' sustainability. Microsoft's Project Natick conducted a two years experiment and found that subsea data centres are feasible, reliable, more energy-efficient, and environmentally sound (Microsoft Research, 2020).

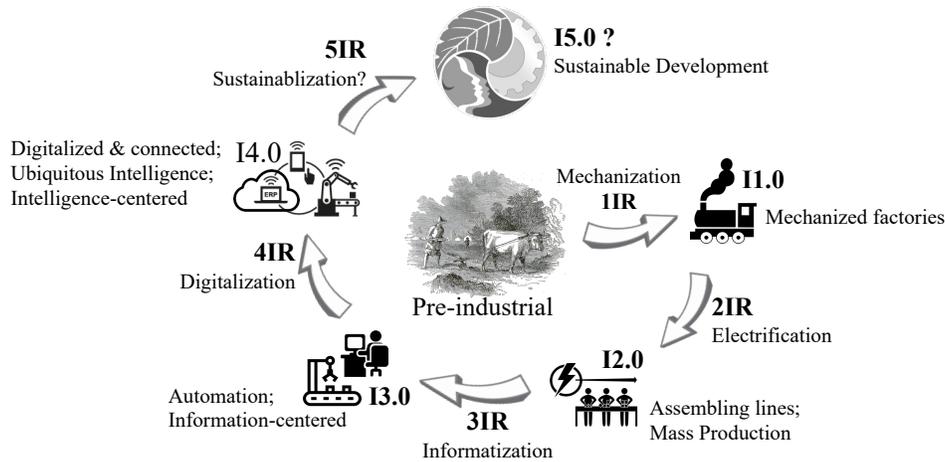

Figure 9. The journey of human society through the industrial revolutions

In a global context, since the first industrial revolution, economic growth has been associated with high environmental costs, including environmental pollution, overexploitation of natural resources, deterioration of ecosystems, loss of biodiversity, and global climate change. The problem is challenging because entities individually lack the motivation to behave environmental-friendly in the market economic mechanism due to environmental externalities. "Sustainable development is development that meets the needs of the present without compromising the ability of future generations to meet their own needs." (UN WCED, 1987) It has become a consensus of the international community (IPCC, 2022; United Nations, 1992). According to the IPCC AR6 report (IPCC, 2022), global warming is speeding up - The increase in global surface temperature in 2011–2020 is 1.09 [0.95 to 1.20] °C higher than in 1850–1900, compared to the increase of 0.19 [0.16 to 0.22] °C in 2003–2012. The trend of increase will reach or exceed 1.5°C in the near term, even for the very low greenhouse gas emissions scenario. Climate change has led to some irreversible impacts on the earth and is approaching a tipping point. This cross-century challenge has become urgent and needs the whole world to act collectively and immediately before it is too late. This global campaign appears geared towards the next industrial revolution for renewable natural resources, particularly renewable energy, reusable materials, and eco-environmental sound economies to achieve harmony between human activities and nature.

In this direction, science, such as disciplines in biological and environmental sciences, energy science, and material science, among others, will be a scientific foundation to reveal facts and provide knowledge about sustainability. Engineering will bring scientific knowledge into the real world by developing technologies and designing innovative solutions for sustainability. Sustainability is a highly complex problem because it is across disciplinary domains, across industry sectors, across regional economies, across cultures, across human groups, across human generations, and across human society and natural systems. The solutions for sustainability need to be coordinated, comprehensive, and systematic. Systems science and engineering will play a unique role in applying systems thinking and designing systematic mechanisms and solutions to address the complex problem of sustainability.

With sustainability in mind, in today's efforts for digital engineering transformation, we should take into account the need for sustainability in digital engineering systems design and development. For example, product provenance can help reuse and recycle after retirement. Particularly, it is critical to consider the sustainability of new digital technologies (Colorado, Velásquez, & Monteiro, 2020; Kellens et al., 2017; Paris, Mokhtarian, Coatanéa, Museau, & Ituarte, 2016). The consideration of sustainability in digitalization will give the emerging digital engineering paradigm(s) a long vision and better address future needs.

## 6.  Concluding Remarks

This article investigated the opportunities and uncertainties in digital engineering transformation and discussed the open issues about the emerging engineering paradigm shifts.

- From investigating the pattern appeared in industrial revolutions, this article revealed that ubiquitous machine intelligence is the defining power brought by the fourth industrial revolution. Digitalization is commonly recognized as the main theme of the 4IR and is also the foundation and necessary condition for the realization of ubiquitous machine intelligence.
- Digital engineering, the digitalization of engineering, is at the core of the fourth industrial revolution. With the broad applications of ubiquitous machine intelligence, many innovative cyber-physical-social smart systems will appear, and traditional engineering systems are also becoming digital smart and connected. The promising opportunities are also accompanied with uncertainties represented by trustworthiness concerns. Given AI as a critical enabling technology in the 4IR, trustworthy AI is a crucial field to ensure systems' trustworthiness in Industry 4.0.
- In the digital smart and connected environment, Digital Engineering needs new concepts, models, tools, methods, theories, methodologies, technologies, and standards. The emerging engineering paradigm shifts include but is beyond the data-intensive paradigm (or data-intensive engineering, the counterpart of data science). Digital engineering transformation towards Industry 4.0 has three essential building blocks: (1) digitalization, (2) leveraging ubiquitous machine intelligence, and (3) building digital trust and security.
- Digitalization is much beyond digitization. Digitization aims to make artifacts machine-readable; digitalization aims to make the things of interest machine-understandable and virtually operatable. Digitalization of a thing of interest should include: (1) digital representation, (2) a unique identifier, (3) metadata including provenance, and (4) a verifiable association of the thing with the above three components.
- The remarkable progress, disruptive impacts, and fast growth of real-world AI applications in the digital and connected environment have triggered concerns and research about the trustworthiness of AI systems. The engineering design community at large is facing an excellent opportunity to bring the new capabilities of ubiquitous machine intelligence and trustworthy AI principles, as well as digital trust, together in various engineering systems design to ensure the trustworthiness of systems in Industry 4.0.

Digital engineering transformation is a crucial process for engineering paradigm shifts in the 4IR. Digital engineering, at the core of the 4IR, is an exciting and broad field to explore and will further lead to "the transformation of entire systems, across (and within) countries, companies, industries and society as a whole" (Schwab, 2017).

## References

AAAI Presidential Panel on Long-Term AI Future. (2009). *Asilomar Study on Long-Term AI Features*. Retrieved from https://www.aaai.org/Organization/asilomar-study.pdf

Adcock, R., Jackson, S., Fairley, D., Singer, J., & Hybertson, D. (2021). Emergence. In *Systems Engineering Body of Knowledge (SEBoK)*. Retrieved from https://www.sebokwiki.org/wiki/Emergence

Ahmed, S., Kim, S., & Wallace, K. M. (2006). A Methodology for Creating Ontologies for Engineering Design. *Journal of Computing and Information Science in Engineering*, *7*(2), 132–140. https://doi.org/10.1115/1.2720879

Armbrust, M., Fox, A., Griffith, R., Joseph, A. D., Katz, R. H., Konwinski, A., … Zaharia, M. (2009). *Above the Clouds: A Berkeley View of Cloud Computing*.

Avizienis, A., Laprie, J.-C., Randell, B., & Landwehr, C. (2004). Basic concepts and taxonomy of dependable and secure computing. *Dependable and Secure Computing, IEEE Transactions On*, *1*(1), 11–33. https://doi.org/10.1109/TDSC.2004.2

Berners-Lee, T., Hall, W., Hendler, J. A., Shadbolt, K. O. and N., & Weitzner, D. J. (2006). A Framework for Web

Science. *Foundations and Trends in Web Science*, *1*(1), 1–130.
Berners-Lee, T., Hendler, J. A., & Lassila, O. (2001). The Semantic Web. *Scientific American*, *284*(5), 35–43.
Black, I. M., Richmond, M., & Kolios, A. (2021). Condition monitoring systems: a systematic literature review on machine-learning methods improving offshore-wind turbine operational management. *International Journal of Sustainable Energy*, *40*(10), 923–946.
Carvalho, T. P., Soares, F. A., Vita, R., Francisco, R. da P., Basto, J. P., & Alcalá, S. G. S. (2019). A systematic literature review of machine learning methods applied to predictive maintenance. *Computers \& Industrial Engineering*, *137*, 106024.
Cheligeer, C., Huang, J., Wu, G., Bhuiyan, N., Xu, Y., & Zeng, Y. (2022). Machine learning in requirements elicitation: a literature review. *Artificial Intelligence for Engineering Design, Analysis and Manufacturing*, *36*(e32), 1–23. https://doi.org/10.1017/S0890060422000166
Chen, D., Doumeingts, G., & Vernadat, F. (2008). Architectures for enterprise integration and interoperability: Past, present and future. *Computers in Industry*, *59*(7), 647–659.
Coatanéa, E., Nagarajan, H., Panicker, S., & Mokhtarian, H. (2022). A prospective analysis of the engineering design discipline evolution based on key influencing trends. *Journal of Integrated Design and Process Science*, *26*(3).
Colorado, H. A., Velásquez, E. I. G., & Monteiro, S. N. (2020). Sustainability of additive manufacturing: the circular economy of materials and environmental perspectives. *Journal of Materials Research and Technology*, *9*(4), 8221–8234. https://doi.org/https://doi.org/10.1016/j.jmrt.2020.04.062
Creswell, A., White, T., Dumoulin, V., Arulkumaran, K., Sengupta, B., & Bharath, A. A. (2018). Generative adversarial networks: An overview. *IEEE Signal Processing Magazine*, *35*(1), 53–65.
Demoly, F., Kim, K.-Y., & Horváth, I. (2019). Ontological engineering for supporting semantic reasoning in design: deriving models based on ontologies for supporting engineering design. *Journal of Engineering Design*, Vol. 30, pp. 405–416. Taylor \& Francis.
Dimassi, S., Demoly, F., Cruz, C., Qi, H. J., Kim, K.-Y., André, J.-C., & Gomes, S. (2021). An ontology-based framework to formalize and represent 4D printing knowledge in design. *Computers in Industry*, *126*, 103374.
Dworschak, F., Dietze, S., Wittmann, M., Schleich, B., & Wartzack, S. (2022). Reinforcement Learning for Engineering Design Automation. *Advanced Engineering Informatics*, *52*, 101612.
EU AI HLEG. (2019). *Ethics guidelines for trustworthy AI*. Retrieved from https://op.europa.eu/en/publication-detail/-/publication/d3988569-0434-11ea-8c1f-01aa75ed71a1
Fox, M. S., Barbuceanu, M., Gruninger, M., & Lin, J. (1998). An Organisation Ontology for Enterprise Modeling. In M. Prietula, K. Carley, & L. Gasser (Eds.), *Simulating Organizations: Computational Models of Institutions and Groups* (pp. 131–152). AAAI/MIT Press.
Fox, M. S., & Gruninger, M. (1998). Enterprise modeling. *AI Magazine*, *19*(3), 109.
Fox, M. S., & Huang, J. (2005). Knowledge Provenance in Enterprise Information. *Int. J. of Production Research*, *43*(20), 4471–4492.
Future of Life Institute. (2017). Asilomar AI Principles. Retrieved December 2, 2021, from https://futureoflife.org/ai-principles/
Goodfellow, I., Pouget-Abadie, J., Mirza, M., Xu, B., Warde-Farley, D., Ozair, S., … Bengio, Y. (2014). Generative adversarial nets. *Advances in Neural Information Processing Systems*, *27*.
Goodfellow, I., Pouget-Abadie, J., Mirza, M., Xu, B., Warde-Farley, D., Ozair, S., … Bengio, Y. (2020). Generative adversarial networks. *Communications of the ACM*, *63*(11), 139–144.
Goranson, H. T. (2002). ICEIMT: History and Challenges. In K. Kosanke, R. Jochem, G. Neil, James, & A. O. Bas (Eds.), *Enterprise Inter- and Intra-Organizational Integration - IFIP TC5/WG5.12 International Conference on Enterprise Integration and Modeling Technology* (pp. 7–14). Kluwer Academic Publishers.
Gray, J. (2007). *eScience -- A Transformed Scientific Method*. Retrieved from http://research.microsoft.com/en-us/um/people/gray/talks/NRC-CSTB_eScience.ppt.
Gruninger, M., & Fox, M. S. (1995). Methodology for the Design and Evaluation of Ontologies. *Workshop on Basic Ontological Issues in Knowledge Sharing, IJCAI-1995*.
Hey, T., Tansley, S., & Tolle, K. M. (2009). *The fourth paradigm: data-intensive scientific discovery*. Microsoft Research.
Hilbert, M., & López, P. (2011). The world's technological capacity to store, communicate, and compute information. *Science*, *332*(6025), 60–65.
Horváth, I. (2022). Designing next-generation cyber-physical systems: Why is it an issue? *Journal of Integrated Design and Process Science*, (Preprint), 1–33.
Huang, J. (2017). Building Intelligence in Digital Transformation. *SDPS Transactions: Journal of Integrated Design*

and Process Science*, *21*(4), 1–4.

Huang, J. (2018). From big data to knowledge: Issues of provenance, trust, and scientific computing integrity. *2018 IEEE International Conference on Big Data (Big Data)*, 2197–2205.

Huang, J., Beling, P., Freeman, L., & Zeng, Y. (2021). Trustworthy AI for Digital Engineering Transformation. *SDPS Transactions: Journal of Integrated Design and Process Science*, *25*(1), 1--7.

Huang, J., & Fox, M. S. (2006). An Ontology of Trust -- Formal Semantics and Transitivity. *Proceedings of the 8th International Conference on Electronic Commerce*, 259–270. ACM.

Huang, J., Gheorghe, A., Handley, H., Pazos, P., Pinto, A., Kovacic, S., … Daniels, C. (2020). Towards digital engineering: the advent of digital systems engineering. *Int. J. System of Systems Engineering*, *10*(3), 234–261. https://doi.org/10.1504/IJSSE.2020.109737

Huang, J., & Nicol, D. M. (2013). Trust mechanisms for cloud computing. *Journal of Cloud Computing*, *2*(1). Retrieved from http://www.journalofcloudcomputing.com/content/2/1/9

Huang, J., Nicol, D. M., Bobba, R., & Huh, J. H. (2012). A framework integrating attribute-based policies into role-based access control. *Proceedings of the 17th ACM Symposium on Access Control Models and Technologies*, 187–196.

Huang, J., Seck, M. D., & Gheorghe, A. (2016). Towards trustworthy smart cyber-physical-social systems in the era of internet of things. *Proceedings of 11th IEEE System of Systems Engineering Conference (SoSE)*.

IPCC. (2022). Climate change 2022: Impacts, adaptation and vulnerability. In H.-O. Pörtner, D. C. Roberts, H. Adams, C. Adler, P. Aldunce, E. Ali, … Others (Eds.), *IPCC Sixth Assessment Report*. https://doi.org/10.1017/9781009325844

ISO/IEC/IEEE. (2010). ISO/IEC/IEEE International Standard - Systems and software engineering -- Vocabulary. *ISO/IEC/IEEE 24765:2010(E)*, pp. 1–418. https://doi.org/10.1109/IEEESTD.2010.5733835

Jin, X., Sandhu, R., & Krishnan, R. (2012). RABAC: role-centric attribute-based access control. *International Conference on Mathematical Methods, Models, and Architectures for Computer Network Security*, 84–96.

Jin, Z., Zhang, Z., Demir, K., & Gu, G. X. (2020). Machine learning for advanced additive manufacturing. *Matter*, *3*(5), 1541–1556.

Kellens, K., Baumers, M., Gutowski, T. G., Flanagan, W., Lifset, R., & Duflou, J. R. (2017). Environmental Dimensions of Additive Manufacturing: Mapping Application Domains and Their Environmental Implications. *Journal of Industrial Ecology*, *21*(S1), S49–S68. https://doi.org/https://doi.org/10.1111/jiec.12629

Kim, H. M., Fox, M. S., & Grüninger, M. (1999). An ontology for quality management—enabling quality problem identification and tracing. *BT Technology Journal*, *17*(4), 131–140.

Kim, K.-Y., Manley, D. G., & Yang, H. (2006). Ontology-based assembly design and information sharing for collaborative product development. *Computer-Aided Design*, *38*(12), 1233–1250.

Kirkpatrick, S., Boniface, M., Bouckaert, S., Grace, P., Jimenez, J., Lahnalampi, T., … Sallstrom, A. (2013). *Future internet research and experimentation: vision and scenarios 2020*. Retrieved from https://eprints.soton.ac.uk/370585/1/370585.pdf

Kuhn, T. (1962). *The structure of scientific revolutions* (2nd Editio). The University of Chicago Press.

Lee, E. A. (2008). Cyber physical systems: Design challenges. *2008 11th IEEE International Symposium on Object and Component-Oriented Real-Time Distributed Computing (ISORC)*, 363–369.

Littman, M. L., Ajunwa, I., Berger, G., Boutilier, C., Currie, M., Doshi-Velez, F., … others. (2021). *Gathering strength, gathering storms: The one hundred year study on artificial intelligence (AI100) 2021 study panel report*. Retrieved from https://ai100.stanford.edu/sites/g/files/sbiybj18871/files/media/file/AI100Report_MT_10.pdf

Liu, H., Wang, Y., Fan, W., Liu, X., Li, Y., Jain, S., … Tang, J. (2021). Trustworthy ai: A computational perspective. *ArXiv Preprint ArXiv:2107.06641*. Retrieved from https://arxiv.org/pdf/2107.06641

Liu, Y., He, D., Obaidat, M. S., Kumar, N., Khan, M. K., & Choo, K.-K. R. (2020). Blockchain-based identity management systems: A review. *Journal of Network and Computer Applications*, *166*, 102731.

Lucas, R., Ang, J., Bergman, K., Borkar, S., Carlson, W., Carrington, L., … others. (2014). *Top ten exascale research challenges*. Retrieved from https://www.osti.gov/servlets/purl/1222713

Luhmann, N. (1979). *Trust and Power*. John Wiley & Sons Ltd.

Madni, A. M., & Sievers, M. (2018). Model-based systems engineering: Motivation, current status, and research opportunities. *Systems Engineering*, *21*(3), 172–190.

Mayer, R. C., Davis, J. H., & Schoorman, F. D. (1995). An Integrative Model of Organizational Trust: Past, Present, and Future. *Academic of Management Review*, *20*(3), 709–734.

Mehrabi, N., Morstatter, F., Saxena, N., Lerman, K., & Galstyan, A. (2021). A survey on bias and fairness in

22    *J Huang / Digital engineering transformation with trustworthy AI towards Industry 4.0: emerging paradigm shifts*machine learning. *ACM Computing Surveys (CSUR)*, *54*(6), 1–35.
Meng, L., McWilliams, B., Jarosinski, W., Park, H.-Y., Jung, Y.-G., Lee, J., & Zhang, J. (2020). Machine learning in additive manufacturing: a review. *Jom*, *72*(6), 2363–2377.
Microsoft Research. (2020). Project Natick. Retrieved October 28, 2022, from https://natick.research.microsoft.com/
Mokammel, F., Coatanéa, E., Coatanéa, J., Nenchev, V., Blanco, E., & Pietola, M. (2018). Automatic requirements extraction, analysis, and graph representation using an approach derived from computational linguistics. *Systems Engineering*, *21*(6), 555–575.
NIST. (2011). *NIST Cloud Computing Standards Roadmap, NIST CCSRWG-092, First Edition*. NIST.
NIST Big Data Public Working Group. (2019). NIST Big Data Interoperability Framework: Volume 1, Definitions Version 3. In *NIST Special Publication 1500-1r2*. Retrieved from https://doi.org/10.6028/NIST.SP.1500-1r2
NSF. (2008). NSF 08-538: Computer Systems Research (CSR). Retrieved from NSF website: https://www.nsf.gov/pubs/2008/nsf08538/nsf08538.pdf
NSF. (2019). National Artificial Intelligence (AI) Research Institutes Accelerating Research, Transforming Society, and Growing the American Workforce (Program Solicitation, NSF 20-503). Retrieved January 1, 2020, from NSF website: https://www.nsf.gov/pubs/2020/nsf20503/nsf20503.htm
Pan, J., Huang, J., Cheng, G., & Zeng, Y. (2022). Reinforcement learning for automatic quadrilateral mesh generation: a soft actor-critic approach. *Neural Networks*, *157*, 288–304. Retrieved from https://arxiv.org/abs/2203.11203
Paris, H., Mokhtarian, H., Coatanéa, E., Museau, M., & Ituarte, I. F. (2016). Comparative environmental impacts of additive and subtractive manufacturing technologies. *CIRP Annals*, *65*(1), 29–32. https://doi.org/https://doi.org/10.1016/j.cirp.2016.04.036
Peisert, S., Cybenko, G., & Jajodia, S. (2015). *ASCR Cybersecurity for Scientific Computing Integrity, DOE Workshop Report, January 7-9, 2015, Rockville, MD*. Retrieved from https://crd.lbl.gov/assets/Uploads/Final-ASCR-Cybersecurity-ReportR7.pdf
Rajkumar, R., Lee, I., Sha, L., & Stankovic, J. (2010). Cyber-physical systems: the next computing revolution. *Design Automation Conference*, 731–736.
Richards, W. T. (1928). A definition of science. *Journal of Chemical Education*, *5*(7), 874. https://doi.org/10.1021/ed005p874
Schneider, F. B. (1999). *Trust in cyberspace*. National Academies Press.
Schwab, K. (2017). *The fourth industrial revolution*. Retrieved from https://www.google.com/books/edition/The_Fourth_Industrial_Revolution/ST_FDAAAQBAJ?gbpv=1
Servos, D., & Osborn, S. L. (2017). Current research and open problems in attribute-based access control. *ACM Computing Surveys (CSUR)*, *49*(4), 65.
Siddik, M. A. B., Shehabi, A., & Marston, L. (2021). The environmental footprint of data centers in the United States. *Environmental Research Letters*, *16*(6), 64017. https://doi.org/10.1088/1748-9326/abfba1
Silver, D., Huang, A., Maddison, C. J., Guez, A., Sifre, L., Van Den Driessche, G., … others. (2016). Mastering the game of Go with deep neural networks and tree search. *Nature*, *529*(7587), 484–489.
Silver, D., Hubert, T., Schrittwieser, J., Antonoglou, I., Lai, M., Guez, A., … others. (2018). A general reinforcement learning algorithm that masters chess, shogi, and Go through self-play. *Science*, *362*(6419), 1140–1144.
Simon, H. A. (1997). *Models of Bounded Rationality* (Vol. 3). The MIT Press.
Sirin, G., Coatanéa, E., Yannou, B., & Landel, E. (2013). *Creating a Domain Ontology to Support the Numerical Models Exchange Between Suppliers and Users in a Complex System Design. Volume 2B*: https://doi.org/10.1115/DETC2013-12266
Smith, R. J. (2022). Engineering. Retrieved October 21, 2022, from Encyclopedia Britannica website: https://www.britannica.com/technology/engineering
Smith, S. (2017). *The Internet of Risky Things: Trusting the devices that surround us*. O'Reilly Media, Inc.
Stone, P., Brooks, R., Brynjolfsson, E., Calo, R., Etzioni, O., Hager, G., … others. (2016). *Artificial intelligence and life in 2030: the one hundred year study on artificial intelligence*. Retrieved from https://ai100.stanford.edu/sites/g/files/sbiybj18871/files/media/file/ai100report10032016fnl_singles.pdf
Su, J., Huang, J., Adams, S., Chang, Q., & Beling, P. A. (2022). Deep multi-agent reinforcement learning for multi-level preventive maintenance in manufacturing systems. *Expert Systems with Applications*, *192*, 116323.
Tavčar, J., & Horvath, I. (2018). A review of the principles of designing smart cyber-physical systems for run-time adaptation: Learned lessons and open issues. *IEEE Transactions on Systems, Man, and Cybernetics: Systems*, *49*(1), 145–158.
UN WCED. (1987). *Our Common Future - Report of World Commission on Environment and Development* (G. H.

Brundtland, Ed.). Retrieved from https://sustainabledevelopment.un.org/content/documents/5987our-common-future.pdf

United Nations. (1992). *Agenda 21*.

US DoD. (2018, June). *Digital Engineering Strategy*. Retrieved from https://fas.org/man/eprint/digeng-2018.pdf

Vernadat, F. (2020). Enterprise modelling: Research review and outlook. *Computers in Industry*, *122*, 103265.

Vigil, M., Cabarcas, D., Buchmann, J., & Huang, J. (2013). Assessing trust in the long-term protection of documents. *2013 IEEE Symposium on Computers and Communications (ISCC)*, 185–191.

Walterbusch, M., Gräuler, M., & Teuteberg, F. (2014). *How trust is defined: A qualitative and quantitative analysis of scientific literature*.

Wang, C., Tan, X. P., Tor, S. B., & Lim, C. S. (2020). Machine learning in additive manufacturing: State-of-the-art and perspectives. *Additive Manufacturing*, *36*, 101538.

Waymo One. (2020). Waymo is opening its fully driverless service to the general public in Phoenix. Retrieved January 1, 2022, from blog.waymo.com website: https://blog.waymo.com/2020/10/waymo-is-opening-its-fully-driverless.html

Weiser, M. (1996). Open House. *ITP Review, NYU*. Retrieved from https://web.archive.org/web/20160310200237/http://www.ubiq.com/hypertext/weiser/wholehouse.doc

Wing, J. M. (2006). Computational thinking. *Communications of the ACM*, *49*(3), 33–35.

Wing, J. M. (2012). Computational Thinking. *Microsoft Research Asia Faculty Summit 2012*. Retrieved from https://www.microsoft.com/en-us/research/wp-content/uploads/2012/08/Jeannette_Wing.pdf

Wing, J. M. (2020). Trustworthy AI. *ArXiv:2002.06276*. Retrieved from http://arxiv.org/abs/2002.06276

Zeng, Y. (2004). Environment-based formulation of design problem. *Journal of Integrated Design and Process Science*, *8*(4), 45–63.

Zeng, Y. (2015). Environment-based design (EBD): A methodology for transdisciplinary design+. *Journal of Integrated Design and Process Science*, *19*(1), 5–24.

Zeng, Y. (2020). Environment: The First Thing to Look at in Conceptual Design. *Journal of Integrated Design and Process Science*, *24*(1), 45–66.

Zheng, Z., Xie, S., Dai, H.-N., Chen, W., Chen, X., Weng, J., & Imran, M. (2020). An overview on smart contracts: Challenges, advances and platforms. *Future Generation Computer Systems*, *105*, 475–491. https://doi.org/https://doi.org/10.1016/j.future.2019.12.019

Zimmerman, P., Gilbert, T., & Salvatore, F. (2019). Digital engineering transformation across the Department of Defense. *The Journal of Defense Modeling and Simulation*, *16*(4), 325–338.